\documentclass[a4paper,11pt]{article}
\pdfoutput=1 

\usepackage{amsmath}
\usepackage{amsfonts}
\usepackage{authblk}
\usepackage{graphicx}

\def\calR{\mathcal{R}}
\def\E{\mathbb{E}}

\def\al{\alpha}
\def\be{\beta}
\def\ga{\gamma}
\def\ep{\epsilon}
\def\de{\delta}
\def\vp{\varphi}
\def\ka{\kappa}
\def\si{\sigma}
\def\te{\theta}

\def\Om{\Omega}
\def\La{\Lambda}
\def\Ga{\Gamma}
\def\na{\nabla}
\def\<{\kern-1pt}

\def\Frac[#1/#2]{\frac{#1}{#2}}
\def\({\left(}	\def\){\right)}
\def\[{\left[}	\def\]{\right]}
\def\LaCDM{$\La$CDM}

\title{\boldmath Hubble drift in Palatini $f(\calR)$-theories}

\author[a]{L.Del Vecchio\thanks{leonardo.delvecchio@edu.unito.it}}
\author[a, b]{L.Fatibene\thanks{lorenzo.fatibene@unito.it}}
\author[c,d,e]{S.Capozziello\thanks{capozzie@na.infn.it}}
\author[a]{M.Ferraris\thanks{marco.ferraris@unito.it}}
\author[f]{P.Pinto\thanks{paolopinto91@gmail.com}}
\author[b,g,h]{S.Camera\thanks{stefano.camera@unito.it}}

\affil[a]{\small Dipartimento di  Matematica, University of Torino,\break
via Carlo Alberto 10, 10123 Torino (Italy)}
\affil[b]{\small Istituto Nazionale di Fisica Nucleare (INFN), Sezione di Torino, \break via P. Giuria 1, 10125 Torino, (Italy)}
\affil[c]{Dipartimento di Fisica, University of Napoli {\it``Federico II''} \break
via Cinthia, I-80126, Napoli (Italy)}
\affil[d]{Istituto Nazionale di Fisica Nucleare (INFN), Sez. di Napoli, \break
via Cinthia 9, I-80126 Napoli (Italy)}
\affil[e]{Gran Sasso Science Institute (GSSI), \break
via F. Crispi 7, I-67100, L'Aquila (Italy)}
\affil[f]{Physics Department, Lancaster University, Lancaster LA1 4YB (UK)}
\affil[g]{Dipartimento di Fisica, University of Torino, \break via P. Giuria 1, 10125 Torino, (Italy)}
\affil[h]{INAF, Osservatorio Astrofisico di Torino, \break strada osservatorio 20, 10025 Pino Torinese (Italy)}

\begin{document}
\maketitle

\begin{abstract}
In a Palatini $f(\calR)$-model, we define chronodynamical effects due to the choice of atomic clocks as standard reference clocks 
and we develop a formalism able to quantitatively separate them from the usual effective dark sources one has in extended theories,
{namely the ones obtained by recasting field equations for $\tilde g$ in the form of Einstein equations}.

We apply the formalism to Hubble drift and briefly discuss the issue about the physical frame.
In particular, we  {shall} argue that there is no {one single}  physical frame, {for example, in the sense one 
defines measure in one frame while test particles goes along geodesics in the other frame}. 
That is the physical characteristic of extended gravity.
As an example, we discuss how Jordan frame may be well suited to discuss cosmology, though it fails within the solar system.

\end{abstract}

\medskip 

\flushbottom

\section{Introduction}
\label{sec:intro}

Cosmology provides today good quality data (see \cite{CMB}, \cite{WMap}, \cite{PlanckExp}, \cite{FermiExp}) to test gravitational theories.
All in all, the universe is the biggest-longest standing experiment we have access to.

Standard GR has proven extremely well in vacuum and stellar scales (within solar system, binary systems, BH coalescence; see \cite{Virgo}).
Unfortunately, one cannot measure all quantities of interest; if not for other reasons, cosmology is practically a coarse graining, even just for that.
For example, we cannot measure directly inertial masses of a galaxy without relying on the model.

Gravity is considerably less well understood and tested in non-vacuum solutions.
Whenever we consider non-vacuum solutions  (galaxies and cosmology; see \cite{ClusterDM1}, \cite{ClusterDM2}, \cite{GalaxyDM1}, \cite{GalaxyDM2}, \cite{Lensing}, \cite{DarkSectiorSupernovae}) we see effects which cannot be accounted by standard GR {\it and} the gravitational sources we see.

One possible solution is to add {\it dark sources}, i.e.~{fundamental fields} which act as gravitational sources though cannot be seen
directly and have no effect other than the gravitational ones, allegedly because they do not interact with the electromagnetic field. That is the \LaCDM{} model approach.
The result is that visible sources correspond to about $4\%$ of all sources. 
According to this model, most of the sources of gravitational field are invisible, about $25\%$ is some sort of WIMP {\it ``exotic''} matter, about $71\%$ some sort of dark energy,
specifically in the form of a cosmological constant.
The dark contributions are approximately invisible at small scales as in the solar system.

Another approach is the modification of gravitational interaction to account for the effects, hopefully without introducing extra sources.
In these approaches, there is no extra sources, the effects are due to modified gravitational dynamics regarded as {\it effective sources}.

Some modified gravitational models (MoND, conformal gravity; see \cite{MOND}, \cite{CG1}, \cite{CG2}, \cite{CG3}) are paradigm shift with respect to standard GR. 
Of course, they are required to locally account for standard GR, by a sort of natural selection. 
In other cases, new models {\it extend} standard GR which is contained as a degenerate particular case; see \cite{Review}.

An example of {\it extended gravity} are Palatini $f(\calR)$-theories in which one considers a family of (regular) functions $f(\calR)$
so that it contains the special case $f(\calR)= \calR$; see \cite{ETG}, \cite{reviewPalatinif(R)}, \cite{OlmoNostro}, \cite{Olmo2011}, \cite{Borowiec}, \cite{ETC}, \cite{EqBD},
\cite{no-go}, \cite{Olmo}, \cite{Mana}, \cite{Wojnar}, \cite{Pinto1}.
That is a relevant structure, since it allows to {\it switch on} the effects starting from standard GR and observe the behaviour of observables.
It also gives a canonical meaning to best fit procedures which are hence performed within the model family to check which model in the family (corresponding to some values of {\it model parameters}) best fits with observations.
Similar, not exactly dynamically equivalent, models are the purely metric $f(R)$-theories; see \cite{Od1}, \cite{Od2}, \cite{Od3}, \cite{Roshan},
\cite{Ca1}, \cite{Ca2}, \cite{Ca3}.

The Palatini $f(\calR)$-theories have some extra bonus characteristics. They automatically implement Ehlers-Pirani-Schild (EPS) framework (see \cite{EPS}, \cite{EPSNostro}, \cite{book2}, \cite{Perlick}), in which geometry of spacetime 
{has the structure of} a Weyl geometry, not {of a Lorentzian one}.
That means that the fundamental fields are a Lorentzian metric $g$ (to describe distances and mediate matter coupling),  a (torsionless) connection $\tilde \Ga$
(to describe free fall), and some set of (tensor) matter fields $\psi$.

The dynamical connection $\tilde \Ga$ turns out to be metric on solutions, though not the Levi-Civita connection $\{g\}$ but { that} of a different conformal metric $\tilde g= \vp \cdot g$.
The scalar field $\vp$ is called the {\it conformal factor}.

 The conformal factor depends on the function $f(\calR)$ one chooses for the model. It is not an extra degree of freedom being a function of the metric field $g$ (and matter).
 Since one has functional constraints between $g$, $\vp$, and $\tilde g$, one can {equivalently} recast the action functional and the field equations in terms of $g$ or $\tilde g$.
 One usually assumes that the gravity-matter coupling is expressed through $g$ and $\psi$, alone not by $\tilde g$ or $\vp$.
 When expressed in terms of $\tilde g$ and $\psi$, the matter Lagrangian is then expected to involve the conformal factor as well.
 
Another extra feature of Palatini $f(\calR)$-theories is {\it universality theorem} (see \cite{Universality}) which states that, essentially for any function $f(\calR)$,  the theory in vacuum is dynamically 
equivalent to standard GR with a cosmological constant $\La$ which has values determined by the function $f(\calR)$.
This is important since it recovers the behaviour of standard GR in vacuum, at least whenever the values of the cosmological constant are small enough.

As we shall briefly review in Section 2, one can recast field equations as Einstein equations for the conformal metric $\tilde g$ and a modified energy-momentum stress tensor $\tilde T_{\mu\nu}$. It contains the energy-momentum tensor $T_{\mu\nu}$ of matter Lagrangian, which is called the {\it visible matter} stress tensor, 
as well as effective contributions which are expressed in terms of the function $f(\calR)$ (which, of course, disappear for $f(\calR)=\calR$).
The extra sources are called {\it effective (dark) sources}, meaning that the model $f(\calR)$ should be chosen to account for non-standard gravitational effects
possibly without requiring extra sources.

In Section $3$, we briefly review cosmological models based on Palatini $f(\calR)$-theories.
We shall also point out there that effective sources {may not be} the only effect introduced by Palatini $f(\calR)$-theories.
In extended gravity, one usually has more than one single metric (e.g.~$g$ and $\tilde g$), as in standard GR.

To be precise, that is not a peculiarity of extended gravity. 
Also in standard GR, every time one has matter, one has a energy-momentum tensor $T_{\mu\nu}$, which can be traced by $g^{\mu\nu}$ to obtain  a scalar field $T$, except special cases as the electromagnetic field for which $T=0$.
Then one can consider a conformal factor $\vp\propto \exp(T)$, thus also in standard GR, beside $g$, one can define infinitely many conformal metrics, e.g.~$\tilde g_\ep= \vp^\ep\cdot g$.
Even if, in this case, the new metric is not naturally encoded in the theory, when we define {\it atomic clocks} as {\it proper clocks}, in principle, they can be chosen to be proper with respect to $g$, $\tilde g_\ep$, as well as any other metric we may define in the model.

Even in standard GR, when the construction is specious, it is a possibility. 
As a matter of fact, there is no absolute definition for a {\it uniform clock} which we would like to use for atomic clocks. We {\it choose} atomic clock to be uniform, by a {\it definition} not on a physical stance.

In a model which naturally has many metrics around, one needs to make an explicit choice to define {\it uniform clocks} to describe atomic clocks; see \cite{Perlick}.
In Palatini $f(\calR)$-theories, also in view of EPS framework, we usually choose atomic clocks to be proper with respect to $g$, not to $\tilde g$.
However, that is still a choice and we would like to be able to { put it to test,} based on experiments instead assuming it.

In Section 4, we set up a formalism able to show that the choice can, in principle, be decided on observations, at least within a family of extended theories.
This formalism also allows  {us} to quantitatively separate the effects due to effective sources from the effects due to the choice of atomic clocks, which are called {\it chronodynamical effects}.
Since we shall show that chronodynamical effects are in principle observable they are not a mathematical {artefact} and theories with different choices of atomic clocks 
are, in principle, observationally distinct.

Finally, in Section 5 we shall present an application to the computation of Hubble drift, i.e.~the relation between the Hubble parameter and the {redshift} (or emission time, or emission distance). The Hubble drift contains information about (and equivalent to) the dynamics of the specific model.
Future surveys (see \cite{Euclid}, {\cite{Eu1}, \cite{Eu2}, \cite{Eu3}, \cite{SKA1}, \cite{SKA2}, \cite{SKA3},} \cite{HD}) promise to provide data about that which hence in principle can be used to test models.

\section{Notation and Palatini $f(\calR)$-theories}

Let us consider a spacetime $M$ of dimension $m=4$, with  a Lorentzian metric $g$, a (torsionless) connection $\tilde \Ga$, and a set of matter fields $\psi$ as fundamental fields.
Let us set $\calR :=g^{\mu\nu} \tilde R_{\mu\nu}$ and $ \tilde R_{\mu\nu}$ is the Ricci tensor of the connection $\tilde \Ga$.

Then the action  functional is given as
\begin{equation}
A_D(g, \tilde \Ga, \psi)= \int_D \(\frac{\sqrt{g}}{2\ka} f(\calR) + L_m(g, \psi)\) d\si
\qquad\qquad
\ka:=  \frac{8\pi G}{c^3}
\label{Action}
\end{equation}
where $\sqrt{g}$ is the usual square root of the absolute value of the determinant of the metric tensor 
and $\sqrt{g}d\si$ is the volume element induced by the metric $g$.
The function  $f(\calR)$ is a generic function (except few degenerate cases), for example a function which is almost anywhere analytic.

Field equations for the action (\ref{Action}) are obtained by varying with respect to $\de g^{\mu\nu}$, $\de \tilde \Ga^\al_{\be\nu}$, and $\de \psi$:
\begin{equation}
\begin{cases}
 f'(\calR) \tilde R_{\mu\nu} -\Frac[1/2] f(\calR)g_{\mu\nu} =\ka T_{\mu\nu} \\
\tilde\na_\al (\sqrt{g} f'(\calR) g^{\be\mu})=0\\
\E =0\\
\end{cases}
\label{JFE}
\end{equation}
In general, the second {equation} is solved by defining a conformal factor $\vp= f'(\calR)$,  a conformal metric $\tilde g_{\mu\nu}= \vp g_{\mu\nu}$
and by showing that $\tilde \Ga=\{\tilde g\}$ is thence the general solution of the second field equation (which, written in terms of $\tilde g$ and $\tilde \Ga$,
is actually algebraic, in fact linear, in $\tilde \Ga$).

The third equation $\E =0$ is obtained as variation of the action with respect to the matter fields $\psi$. It describes how matter fields evolve in the gravitational field.
On the other hand, $T_{\mu\nu}$ depends on $\psi$ so that matter fields are sources for the gravitational field.

By tracing the first equation by means of $g^{\mu\nu}$, one obtains the so-called {\it master equation}
\begin{equation}
f'(\calR) \calR -2f(\calR) =\ka T
\end{equation}
where we set $T:= g^{\mu\nu}T_{\mu\nu}$.
This is also an algebraic equation in $\calR$ and $T$ which generically can be (at least locally) solved for $\calR= \calR(T)$, so that the curvature $\calR$ along solutions can be expressed as a (model dependent but) fixed function of the matter content $T$.

In vacuum, one has $T=0$ and the zeros of the master equation define constant values the curvature can {assume}; see \cite{Universality}.
We say that the function $f(\calR)$ is {\it regular} if the master equation has simple isolated zeros or, more generally, if it can be inverted in branches almost anywhere.

At this point, the first field equation can be recast as the Einstein equation for the metric $\tilde g$ 
\begin{equation}
\tilde R_{\mu\nu} -\Frac[1/2] \tilde R \tilde g_{\mu\nu} = \ka \tilde T_{\mu\nu} 
\label{EE}
\end{equation}
where we introduce the {\it scalar curvature $\tilde R= \tilde g^{\mu\nu}\tilde R_{\mu\nu}= \vp^{-1}\calR$ of the conformal metric $\tilde g$} and the {\it effective energy--momentum stress tensor}
\begin{equation}
\tilde T_{\mu\nu} := \Frac[1/f'(\calR)] \( T_{\mu\nu}  - \Frac[ f'(\calR) \calR - f(\calR) /2\ka]  g_{\mu\nu}\)
\end{equation}
and the scalar curvature $\tilde R$ of the conformal metric $\tilde g$.
Working with $\tilde g$ only is often called the {\it Einstein frame}, while using $g$ is called the {\it Jordan frame}.

Let us notice that the Einstein frame is often characterised as the choice in which field equations are in Einstein form, though with a modified source term, 
as opposed to (\ref{JFE}). However, this is of course false (and meaningless). Since equations  (\ref{JFE}) are linear in the Ricci tensor $\tilde R_{\mu\nu}$ 
and the metrics are conformal to each other, one can recast those equations in Einstein form as well, at the price of sending on the right hand side the extra terms expressed as function of matter fields by using the master equation.

{\it If one wants to justify $g$ and $\tilde g$, in fact, one needs to say that $g$ is minimally coupled to matter, while $\tilde g$ is not.}

The effective energy--momentum stress tensor $\tilde T_{\mu\nu}$ differs from the original $T_{\mu\nu}$
which is the usual variation of the matter Lagrangian with respect to the metric $\de g^{\mu\nu}$. 
Whatever {\it visible matter} is, it is described by $T_{\mu\nu}$, then $\tilde T_{\mu\nu}$ directly gets extra contributions from the modified dynamics, i.e.~from the function $f(\calR)$
which, hopefully, by choosing it accordingly, can be used to model dark matter and energy as effective sources.
The appearance of dark sources as effective sources is a consequence of the modification of dynamics and it is accordingly called a {\it dynamical effect}.

This is not the only effect in extended theories. Also the odd definition of atomic clocks (which are free falling with respect to $\tilde g$ but proper with respect to $g$) 
produces extra accelerations in particles. These accelerations are universal, i.e.~they are easily confused with an extra gravitational field acting on all test particles equally which, when reviewed in a standard GR setting, calls for other sources. 
These are called {\it chronodynamical effects}.
Hereafter, we shall investigate the combination of these two types of effects in cosmology, we show that both have observational consequences, and that we can separate them from dynamical effects.

\section{Cosmologies based on Palatini $f(\calR)$-theories}

In order to extend cosmological principle to an (integrable) Weyl geometry $(M, g, \{\tilde g\})$, we should declare whether we ask $g$ or $\tilde g$ to be spatially homogeneous and isotropic.
However, since the master equation holds true on solutions, the conformal factor is a function of $t$ only and $g$ is spatially homogeneous and isotropic iff $\tilde g$ is.
Only, if $g$ is in FLRW form in coordinate $(t, r, \te,\phi)$, with a scale factor $a$, then $\tilde g$ is in FLRW form in coordinate $(\tilde t, r, \te,\phi)$, with a scale factor $\tilde a= \sqrt{\vp}\> a$.
The conformal factor is a function only of time and the new time is defined by integrating $d\tilde t = \sqrt{\vp} \> dt$.

Thus one has a Friedmann equation both for $a$ and $\tilde a$
\begin{equation}
\dot a^2 = \Phi(a)
\qquad\qquad
\dot {\tilde a}^2 = \tilde \Phi(\tilde a)
\label{FE}
\end{equation}
which are, of course, defined to be equivalent.
{ The Friedmann equation is in the form of a Weierstrass equation, so we can qualitatively study the evolution of the scale factor.
The notation for Weierstrass equation r.h.s. as $\Phi(a)$ is standard in mathematical literature. It has not to be confused with 
Bardeen gauge invariant metric potentials.}
The specific form of the function $\tilde \Phi(\tilde a)$   is obtained by expanding the Einstein equations (\ref{EE}). 
The function $\Phi(a)$ is instead obtained by using the relation $\tilde a= \sqrt{\vp(a)}\> a$ as
\begin{equation}
\Phi(a) :=   \vp(a)  \({\Frac[d\tilde a/d a]}\)^{-2} \tilde \Phi (\tilde  a) 
\qquad\qquad\hbox{where }\quad
\Frac[d\tilde a/d a] := \Frac[ 2\vp +{a\Frac[d\vp/da](a)}/ 2\sqrt{\vp}]
\end{equation}

As a consequence of the cosmological principle {at the level of the Universe's background evolution}, the energy--momentum tensor $T_{\mu\nu}$ is in the form of a perfect fluid energy-momentum tensor, namely
\begin{equation}
T_{\mu\nu} = c^{-1} \( (\rho c^2+p) u_\mu u_\nu + p g_{\mu\nu}\)
\end{equation}
for some time-like, future directed, $g$-unit, comoving vector $u^\mu$. 
Also $\tilde T_{\mu\nu}$ is a perfect fluid energy-momentum 
tensor using $\tilde g$ and a suitable $\tilde g$-unit vector $\tilde u$ as well as different effective pressure and density $\tilde p$ and $\tilde \rho$. 

If we choose a specific $f(\calR)$, e.g.~as in \cite{Pinto1} (where also issues about regularity are discussed), i.e.
\begin{equation}
f(\calR)= \al \calR -\Frac[\be/2] \calR^2 -\Frac[\ga/3] \calR^{-1}
\label{fR}
\end{equation}
the effective density and pressure are given by
\begin{equation}
\tilde \rho =\Frac[  4 \ga\calR -3 \be \calR^4 + 12\ka c \rho\calR^2 / 4\ka c \(3 \al \calR^2-3 \be \calR^3+\ga\) \vp]
\qquad
\tilde p= -\Frac[ 4c \ga\calR -3c \be \calR^4  -12\ka  p\calR^2 / 4\ka  \(3 \al \calR^2-3 \be \calR^3+\ga\) \vp]
\end{equation}
where $\rho$ and $p$ are the total mass density and pressure of visible matter.  

Since $T_{\mu\nu}$ is the variation of a covariant matter Lagrangian, it is conserved, i.e.~$\na_\mu T^{\mu\nu}=0$.
One can show that also the effective energy--momentum stress tensor is conserved though with respect to $\tilde g$, i.e.~$\tilde \na_\mu \tilde T^{\mu\nu}=0$; see \cite{Conservation}.

Conservations are equivalent to the expressions of pressures as $p= -(\rho + \Frac[1/3]a \rho' )$ and $\tilde p= -(\tilde \rho + \Frac[1/3]\tilde a \tilde \rho' )$,
thanks to which only the first Friedmann equations are independent.
Finally, one fixes the EoS for $\rho$, which gives $\rho(a)$ as a function of the scale factor $a$ and the effective EoS for $\tilde \rho(a)$ are obtained.
If we fix dust visible matter, $\rho(a)= \rho_0 a^{-3}$, the effective EoS is non-linear and quite exotic.
It depends on the the function $f(\calR)$ and, in general, it is not even a mixture of polytropic fluids.

We also normalise the scale factors and the conformal factor to be one today (as well as to be always positive).
the idea is to solve the equation $\dot a^2 = \Phi(a)$ for $t(a)$ and to express all quantities in the model as a function of $a$ so to be able to describe all mutual relations; see \cite{Pinto1}.

If we fix the function (\ref{fR}), the master equation reads as
\begin{equation}
\begin{aligned}
 \al \calR-\be \calR^2 &+{\Frac[\ga/3]} \calR^{-1}  - 2 \( \al \calR -\Frac[\be/2] \calR^2 -\Frac[\ga/3] \calR^{-1} \)=\\
 =& - \al \calR  +  \ga \calR^{-1} = \Frac[\ka/c] (3p- \rho c^2) = -\ka c \rho
 \end{aligned}
\end{equation}
which can be solved in two branches (corresponding to the sign of $\calR$) as
\begin{equation}
{}^\pm \calR(a) =  \Frac[ \ka \rho c \pm \sqrt{ \ka^2 \rho^2 c^2+4\al  \ga} /2\al ] 
=  \Frac[ \ka c \rho^d_0  \pm \sqrt{ \ka^2 c^2 \(\rho^d_0\)^2 +4 \al  \ga a^{6}} /2\al a^{3}] 
\label{R(a)}
\end{equation}

In \cite{Pinto1} we used SNIa to best fit parameters. The model is quite degenerate, in particular it needs some more tests to fix $\al$ and $\be$,
so that SNIa then can fix $\ga$.
Here we shall show {the} results for {the} values:
\begin{equation}
\al \simeq 0.095
\qquad\qquad
\be=0.25\> m^2
\qquad\qquad
\ga \simeq  2.463\cdot 10^{-104} \>m^{-4}  
\end{equation}

The conformal factor is to be chosen proportional to $f'(\calR)$
which is everywhere positive if we use ${}^- \calR(a)$, while ${}^+ \calR(a)$ changes sign at about $\rho_1:=1.925\cdot 10^{24} \> kg\> m^{-3}$.
Thus, for the conformal factor to be positive, we need to define it in three branches

\begin{itemize}
\item[-] the branch {A}, with $\calR>0$ and and $\rho\in(\rho_1, +\infty)$  (thus $a\in (0, a_1)$), where the conformal factor is defined as
$\vp_A := -\vp_0 f'(\>{}^+\<\calR)$;

\item[-] the branch B,  with $\calR>0$ and $\rho\in(0,\rho_1)$  (thus $a\in (a_1,+\infty)$), where the conformal factor is defined as
$\vp_B := \vp_0 f'(\>{}^+\<\calR)$;

\item[-] the branch {C}, with $\calR<0$ and $\rho\in(0,+\infty)$ (thus $a\in (0,+\infty)$), where the conformal factor is defined as
$\vp_C := \vp_0 f'(\>{}^-\<\calR)$;

\end{itemize}

\noindent
where $\vp_0$ is a constant to be chosen so that today $\vp(t_0)=1$.

Branch A corresponds to very high densities, so it happened early in the universe. We assume then to currently be on branch $B$ at $a=a_0=1$.
So we choose $\vp_0 := \( f'(\>{}^+\<\calR(a_0=1)) \)^{-1}$.

By using the correct expression for $\vp$ and $\calR$ on each branch, we can compute {the} Friedmann equation
\begin{equation}
\dot {\tilde a}{}^2 = \Frac[\ka c^3/3] \tilde \rho(\tilde a)\> {\tilde a}^2 -k c^2  =: \tilde \Phi (\tilde a)
\end{equation}

In view of the transformation between the two frames induced by the conformal factor, we have
\begin{equation}
\begin{aligned}
\dot a ^2= \Phi(a) := &  \vp(a)  \({\Frac[d\tilde a/d a]}\)^{-2} \tilde \Phi (\tilde \rho(a)) =\\
=&   \vp(a)  \({\Frac[d\tilde a/d a]}\)^{-2}  \(\Frac[{\ka {c^3}}/3]   \tilde \rho(a) \tilde a^2(a)  -kc^2\)
\end{aligned}
\end{equation}

The spatial curvature as a function of the visible matter density $\rho_0$, i.e. 
\begin{equation}
 k(\rho_0) = c^{-2} \(\Frac[{\ka {c^3}}/3]   \tilde \rho(\rho_0) -\omega^2 H_0^2 \)
  \qquad\qquad  
\(\omega :=\Frac[d\tilde a/ d a]{\Big|_{a=1}}\)
 \label{krho}
 \end{equation}
Let us remark that, once the function $f(\calR)$ is fixed, then we know the function $\tilde \rho(\rho)$ and the constant $\omega$.

One can solve the integral
\begin{equation}
t(a) = \int_1^a \Frac[da / \sqrt{\Phi(a)}]
\label{Solution}
\end{equation}
Then the parametric curve ${\lambda}: a \mapsto (t(a), a)$ represents the graph of the function $a(t)$.

The choice of the parameters is done so that the evolution of the scale factor $a(t)$ is quite closed to the prediction of \LaCDM; see \cite{Pinto1}.

\section{Chronodynamical effects}

It has been argued that Palatini $f(\calR)$-theories are just standard GR (for $\tilde g$) in disguise, i.e.~just written with a different choice of fields.
That would be motivated by the fact that the conformal factor is not an independent degree of freedom, but it is determined as a function of the metric and matter.

This is, of course, something which has to be decided by observations. One should pick some quantity which is observable (e.g.~the Hubble drift $H(z)$),
 determine which model theoretic quantity has to be used to predict the result of observations, and test it.

Let us stress that, the choice of the model theoretic quantity is not determined by the action principle, it is an independent choice.
Sometimes, it has been argued that, for example, the action principle does determine the equation for test particles. 
That is only partially true: 
on one hand, it is certainly true that one can obtain an equation as the eikonal  approximation of matter field equation
(so it would only be a matter of knowing how long it takes for the approximation to break down, since we are easily talking about test particles which need to go around for 14By!).
On the other hand, one could rescale the matter field $\psi$ by the conformal factor, define a new matter field $\psi'=\vp^\al \psi$, and obtain a {\it different},
eikonal approximation for test particles of $\psi'$. 
In other words, the eikonal approximation is not invariant with respect to redefinition of fields, which may not be an issue when we have a clear correspondence between fields and the test particles we are considering.
That is often not the case in GR. We do not have a clear and unique description in terms of fields for planets in the solar system, or for an asteroid around a black hole. 
To be honest, we do not even know if a planet is {\it exactly} a test particle: it is made of interacting parts, thus it may be reasonable (rather than not) to assume that one can neglect or average the non-gravitational interactions and regard it, as a whole, as a freely falling test particle.
Hence, we hide all our approximations in the assumption that a given equation describes freely falling test particles (or, equivalently, that a certain field corresponds to test particles), and then test it {\it a posteriori}.
Still that is an independent assumption.

EPS gives us a framework to interprete quantities in a Weyl geometry $(M, g, \{\tilde g\})$: the connection form $\tilde g$ is built explicitly to describe test particles, a representative $g$ of the conformal class $[g]$ is selected to define proper clocks, so atomic clocks can be assumed to measure $g$-proper time.
Of course, the purely Lorenztian geometry $(M, \tilde g, \{\tilde g\})$ is still a possibility.

Once we select the dynamics by choosing the function $f(\calR)$, we have a parametric family of functions (in the case we are considering here depending on the parameters $\boldsymbol{\al}=(\al, \be, \ga)$) which contains standard GR as a special case ($\boldsymbol{\al}_0=(\al=1, \be=0, \ga=0)$ in our case).
We can define observable quantities in the model and require that they reduce to the observables in standard GR, which usually are reasonably well known.
Still, the extension of observable quantities is often not unique.

In \cite{Perlick}, \cite{Pinto1}, \cite{book2}, it has been shown that observation protocols descend from the choice of a model theoretic representation of atomic clocks which are assumed as standard time. EPS provides us with a  clear choice of atomic clocks as clocks which measure $g$-proper time.
Unfortunately, in the limit to standard GR, both the dynamical and chronodynamical effects vanish and we do not have a way to test the two choices, the dynamics $f(\calR)$ and the metric for atomic clocks, separately.

To avoid it, we can introduce a wider Weyl framework, in which atomic clocks (as well as consequently all observational protocols which descend from that)
are proper by some metric $g_s$ which continuously connects $g$ and $\tilde g$. 
As a matter of fact, we are introducing an extra parameter $s$ which selects atomic clocks and allows us to switch on and off the chronodynamical effects in 
Palatini $f(\calR)$-theories, without modifying the dynamics, i.e.~touching $(\al, \be, \ga)$.

Deforming metrics is notoriously difficult, since the space of metrics of a given signature is not affine. However, in a Weyl geometry, metrics are elements in a given conformal class, which instead {\it is} an affine space. In simpler words, we can deform the conformal factor as
\begin{equation}
g_s= \vp_s g
\qquad\qquad
\vp_s = s \vp + (1-s)1
\end{equation} 
so that when $s=0$ one has $g_0=g$ and when $s=1$ one has $g_1=\tilde g$.
Accordingly, we are introducing a new parameter $s\in [0,1]$ so that we can test whether chronodynamical effects are observable
and hence deciding if the atomic clocks are $g$-proper or $\tilde g$-proper on an observational basis instead of assuming it.

\section{Hubble drift}

In \cite{Pinto1} and \cite{book2} we showed that the Hubble parameter we measure, as well as the redshift, are directly related to  
the scale factor of the metric that we use for atomic clocks, i.e.~$g_s$.
Let us denote them by 
\begin{equation}
H_s:= \Frac[1/a_s]\Frac[ da_s / dt_s]= \Phi(a)
\( 1+   {\vp_s}^{-3/2} \Frac[ s  / 2 ]{\Frac[d\vp / da]} \)
\qquad\qquad
z_s:= \Frac[1-a_s/ a_s]
\end{equation}
where $a_s= \sqrt{\vp_s} \>a$ and $dt_s= \sqrt{\vp_s}\> dt$.

Then we can fix $f(\calR)$ as in (\ref{fR}), solve the model and compute $H_s(a)$ and $z_s(a)$.
These functions provide us with a parametric representation of the model theoretic Hubble drift $H(z)$ which should be compared with the observations.
They are functions of the scale factor parameter $a$, as well as of the parameters $\boldsymbol{\al}=(\al, \be, \ga)$ and $s$.

We still do not have data to fit, which will be available with future surveys, see e.g.~\cite{Euclid},
though we can show once and for all that both dynamical and chronodynamical effects are potentially observable.

We first compute and draw (see Figure 1) the Hubble drift $H(z)$ for:

\begin{itemize}
\item[-] {\it standard GR} (blue solid) with $\rho_0^d = 4.89 \cdot 10^{-28} \> kg\>m^{-3}$ (as in \LaCDM),
$\rho_0^r = 10^{-30}\> kg\>m^{-3}$ (as in \LaCDM),
but no cosmological constant. We require $k=0$ and $H_0= 2.373\cdot 10^{-18}\>s^{-1}$.

\item[-] \LaCDM{} (red dash) with $\Om_b = 4.8\%$,
$\Om_r = 10^{-4}$,
$\Om_\La = 70\%$,
$\Om_c \simeq 25$\%,
$k=0$ and $H_0= 2.373\cdot 10^{-18}\>s^{-1}$.

\item[-] Palatini $f(\calR)$-theory (black). For $\ga=\ga_0=2.463\cdot 10^{-104} \>m^{-4}$, $\ga=3\ga_0$, and $\ga= \ga_0/3$.
We set $s=0$ as dictated by EPS and used in \cite{Pinto1}. 

\end{itemize}

\begin{figure}[htbp] 
\hskip-1cm   \hbox{\includegraphics[width=7cm]{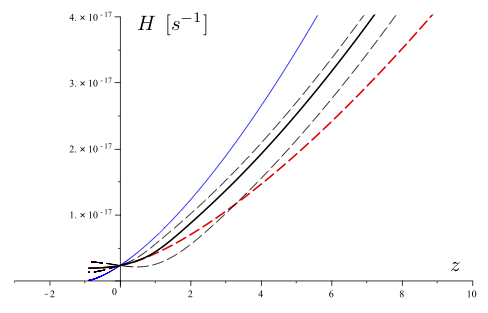}
   \qquad 
   \includegraphics[width=7cm]{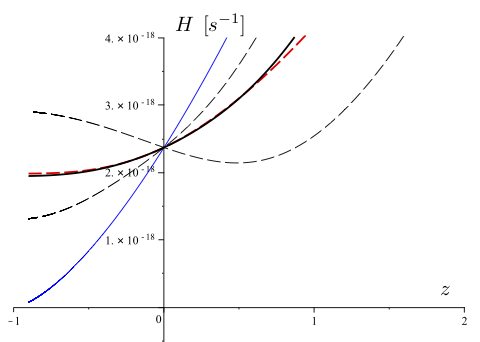}}
   \caption{\small\it the function $H(z)$. Comparison between GR (blue solid), \LaCDM{} (red dash), and Palatini $f(\calR)$-theory (black)
for 3 different values of $\ga$.\hfill\break
a) deep view $z\in [-1, 10]$. \hskip5cm
b) zoom at $z\in [-1, 2]$.
}
   \label{fig:1}
\end{figure}

Analyzing Figure 1 we see that the values of $H(z)$ near today are very sensitive to the value of $\ga$.
For the best fit value obtained in \cite{Pinto1} we have (of course) a good agreement with \LaCDM. Still Palatini $f(\calR)$-theory gives a different prediction for high $z$.

We can see also that the prediction depends on $s$; see Figure 2. We fix $\ga=\ga_0$ and draw the prediction for $s=0$ (black solid), $s=1/2$, $s=1$ (black dotted).

Again chronodynamical effects are small  for small $z$, while they become sensible for $z\sim 6$.
We see that the best fit is obtained for $s=0$, {as expected}.
Let us remark that the functions $H(z)$ have been obtained with no approximation at all.

\begin{figure}[htbp] 
\hskip-1cm   \hbox{\includegraphics[width=7cm]{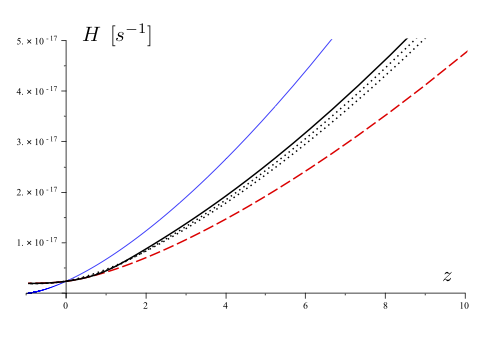}
   \qquad 
   \includegraphics[width=7cm]{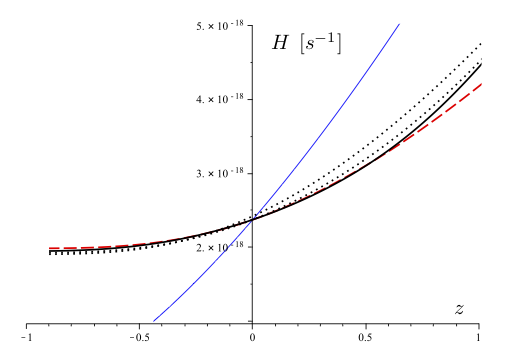}}
   \caption{\small\it The function $H(z)$. Comparison between GR (blue solid), \LaCDM{} (red dash), and Palatini $f(\calR)$-theory (black)
for 3 different values of $s$.\hfill\break
a) deep view $z\in [-1, 10]$. \hskip5cm
b) zoom for $z\in [-1, 1]$.
}
   \label{fig:2}
\end{figure}

\section{Conclusions and Perspectives}

We considered the prediction for the Hubble drift $H(z)$ in a Palatini $f(\calR)$-theory.
In order to keep separate dynamical and chronodynamical effects we introduced an extra 
parameter $s$ which selects the metric which describes atomic clocks. 
For $s=0$ we obtain that atomic clocks are described as proper clocks with respect to $g$,
for $s=1$ we obtain proper clocks with respect to $\tilde g$.

When we consider $s=1$, then $\tilde g$ is responsible for both free fall and metrology.
Accordingly we could reformulate the whole model as a theory for $\tilde g$ which corresponds to
standard GR, an extra matter field (the conformal factor) and non-minimal couplings.

For $s=0$ instead, no metric does everything so the model is {neither} in the Jordan frame {nor} in the Einstein frame.
Let us remark that, as long as we limit to cosmology, and we deal only with the surfaces $t=const$ and the comoving worldlines, 
these are both conformally invariant and we really see no difference between structures in the Jordan and Einstein frames.
Accordingly, we can say we are in the Jordan frame.
Palatini $f(\calR)$-theories precisely model the conjectured effect in which one has, unlike in standard GR, a Weyl geometry on spacetime
to describe gravitational field, in which the metric that describes free fall of test particles is not the metric which one obtains from light rays in vacuum.
As a consequence, the theory in vacuum is dynamically equivalent to standard GR with a cosmological constant, with extra effects which become apparent only in non-vacuum solutions.

One can show that in these cases, there are a number of dynamical equivalence which connects different theories
(namely, non-minimally coupled GR, on one side, Brans-Dicke theory with a potential, on the other side); see \cite{Magnano}, \cite{ETG}. 
However, the dynamical equivalence maps solutions of a theory onto solutions of the other, not necessarily observables into observables
with the same physical interpretation.
For example in Brans-Dicke theory one uses $g$ to describe test particles, while, in the corresponding Palatini $f(\calR)$-theory, $\tilde g$
is used. Hence, for example Mercury, orbits along different trajectories in the two theories; see \cite{MathEquivalence}.
No surprise that if one does not account for that mismatch, Palatini $f(\calR)$-theories may (or may not) pass tests in the solar system.

Often it has been (wrongly) argued that, in view of dynamical equivalence, one can exclude a theory when the other is excluded, which is certainly true if 
the dynamical equivalence extends to a complete physical equivalence, which it is not if there is a mismatch in observables.
It is precisely when we pay attention to solar system tests (or any other situation in which timelike geodesics other than comoving play a role) that we highlight that we are not working in the Jordan frame only.
For this reason, it is particularly important in extended theories of gravitation to have a detailed and rigorous framework for observables which
could eventually account for how a different dynamics can be equivalently seen as a collection of effective dark sources.
That is something to be required to ETG, exactly as one requires to \LaCDM{} a fundamental account for dark sources.

It is certainly true that when one introduces dark sources at a fundamental level then that fundamental sources are expected to have some sort of, maybe tiny, effects other than the gravitational effect.  Hence \LaCDM{} is called to provide evidences of such effects or to find direct evidences of dark sources.

Similarly, extended theories of gravity  are essentially claiming that dark sources are not fundamental, they are produced as an effective counterpart of a modification of dynamics. They need not to explain anything at a fundamental level, they are instead claiming that no direct evidences of dark sources will never be found.
ETG are in principle falsifiable, although one may be ready to trade a bit about partial contributions to the dark sector, unless one believes that we already have an ultimate description of the fundamental level.
However, we should be clear about duties of different approaches; if one claims that dark sources are effective then a detailed and complete account of how observational protocols are deceived into seeing the modified dynamics as a dark source needs to be provided.
In other words, if people who call for exotic sources must provide evidences, other than the gravitational ones, for such extra sources, then, on the other hand, people in ETG must explicitly account for how observational protocols extend to their models and how they account for observations without dark sources.

\medskip
In this paper, we considered an observable in cosmology, namely the Hubble drift $H(z)$, and analyse it in the cosmology based on a Palatini $f(\calR)$-theory.
We showed that we can disentangle dynamical and chronodynamical effects and that the Hubble drift is both affected by changes of the parameters $\boldsymbol{\al}=(\al, \be, \ga)$ describing the dynamics, and by changes of the parameter $s$ describing the chronodynamical effects.
Both the effects are in principle observable within the range of most distant objects we see in the universe, i.e.~$z\in [-1, 10]$.
These are deviation of that specific Palatini $f(\calR)$-theory with respect to \LaCDM{} which can be used to disprove either of the two.

We have to stress that, in the Palatini $f(\calR)$-model we considered, though that is true in general, there is a lot of degeneracy. 
While it is easy to show that two particular models in the family, for different values of the parameters, are observationally different, 
there are families of parameters which fits a particular test. Accordingly, one needs more tests to remove degeneracy and then the model is ready to make real predictions. 
In view of this degeneracy, it is pretty obvious that, adding parameters to a cosmological model, one is almost certain to be able to fit a specific observation,
and that is a relatively nonsensical game.
The real challenge is to be able to explain more effects than one uses to fit the theory's parameters and, in order to do that, one needs a satisfactory control on observables in the theory, which {allows for a determination of} which model-theoretic quantity corresponds to a given observable we measure.
We did it in \cite{Pinto1}, where we reviewed the standard GR argument and showed that the observed Hubble parameter is the one connected to the metric which is used to describe atomic clocks, and more generally, that all basic observational protocols are derived from clocks. 
Unfortunately, since all protocols in relativistic theories are model--dependent, that is a huge effort required in general.

Palatini $f(\calR)$-theories are, in that respect, an extremely conservative approach among modified gravity. They directly implement EPS framework,
they contain standard GR as a special case, so one can deform models continuously, one into the other, and compare them rigorously.
Still, in principle, they introduce with the function $f(\calR)$ infinitely many parameters so that, if the meaning of a specific model depending on a finite number of parameters is clear, not the same can be said for a generic $f(\calR)$-theory.
There are simply no result (other than the universality theorem) or framework to deal with a generic ETG and this appears currently out of reach.

{\small
 \section*{Acknowledgments}
This article is based upon work from COST Action (CA15117 CANTATA), supported by COST (European Cooperation in Science and Technology).
We acknowledge  the contribution of INFN (IS-QGSKY), the local research project {\it Metodi Geometrici in Fisica Matematica e Applicazioni (2017)} of Dipartimento di Matematica of University of Torino (Italy). 
SC is supported by the Italian Ministry of Education, University and Research (MIUR) through Rita Levi Montalcini project `\textsc{prometheus} -- Probing and Relating Observables with Multi-wavelength Experiments To Help Enlightening the Universe's Structure', and by the `Departments of Excellence 2018-2022' Grant awarded by MIUR (L.~232/2016).
This paper is also supported by INdAM-GNFM.

}

\end{document}